\documentclass[a4paper,12pt]{article}
\usepackage[applemac]{inputenc}
\usepackage{amssymb,amsmath}
\usepackage{graphicx}
\usepackage[english]{babel}
\usepackage{hyperref}

\newcommand{\kB}{k_{\mathrm{B}}}

\begin{document}
\begin{titlepage}
\title{Carnot cycle for an oscillator}

\author{ Jacques \textsc{Arnaud} 
\thanks{Mas Liron, F30440 Saint Martial, France},
Laurent \textsc{Chusseau}
\thanks{Centre d'\'Electronique et de Micro-opto\'electronique de
Montpellier, Unit\'e Mixte de Recherche n°5507 au CNRS, Universit\'e
Montpellier II, F34095 Montpellier, France},
Fabrice \textsc{Philippe}
\thanks{D\'epartement de Math\'ematiques et Informatique Appliqu\'ees,
Universit\'e Paul Val\'ery, F34199 Montpellier, France.  Also with
LIRMM, 161 rue Ada, F34392 Montpellier, France.} }

\maketitle

\begin{abstract}      
Carnot established in 1824 that the efficiency of cyclic engines
operating between a hot bath at absolute temperature $T_{hot}$ and a
bath at a lower temperature $T_{cold}$ cannot exceed
$1-T_{cold}/T_{hot}$.  We show that linear oscillators alternately in
contact with hot and cold baths obey this principle in the quantum as
well as in the classical regime.  The expression of the work performed
is derived from a simple prescription.  Reversible and non-reversible
cycles are illustrated.  The paper begins with historical
considerations and is essentially self-contained.
\end{abstract}

\end{titlepage}

\section{Introduction}

The purpose of this paper is to apply Carnot cycles to linear
oscillators in the quantum regime rather than to gas-filled cylinders
as is done in most Thermodynamics text-books.  Of course the forces
involved in such systems are tiny at usual temperatures.  But it is
instructive to verify for such a simple model that the average work
performed per cycle is accurately given by the Carnot principle. 
Because the system is small the work performed may fluctuate
significantly from cycle to cycle.  We thus distinguish deterministic
(italic letters) and fluctuating (roman letters) quantities, even
though we are presently interested mostly in average values.

The paper is essentially self-contained.  It is hoped that the readers
will find useful our concise presentation and illustration of the
relevant laws of Thermodynamics, Statistical Mechanics, and Quantum
Theory.  Quantities that are not strictly required have been set
aside, in particular \emph{system} entropies and temperatures.

Let us first attempt to summarize the subtle reasonings that enabled
Sadi Carnot early in the 19th century to prove that the maximum
efficiency of any heat engine is given by the formula
$\eta_{C}=1-T_{cold}/T_{hot}$, where $T_{cold},T_{hot}$ are the
absolute temperatures of the cold and hot heat reservoirs,
respectively.  This achievement was made with few empirical results
available.  Indeed, as Carnot calculated it, the efficiencies of heat
engines fabricated at the time were, at best, only 5$\%$ of the
maximum efficiency $\eta_{C}$, so that observations did not provide
any hint to the value of the maximum efficiency attainable.  It was
well known, however, that heat never flows from cold to hot bodies
spontaneously.  Only heat pumps, which require a supply of mechanical
or electrical energy, may reverse the natural heat-flow direction.  In
the above formula, the efficiency is defined as the ratio of the
average work $\Delta W$ performed by the heat engine per cycle, for
example through the lifting of a weight, and the upper reservoir
average energy loss $-\Delta Q_{hot}$\footnote{The minus sign is
introduced for later convenience: amounts of heat are defined as
positive when they are \emph{added} to the baths.  Likewise, entropies
are defined as positive when they are \emph{produced}.}.  The hot
reservoir may consist of the liquid and vapor phases of a substance in
a state of equilibrium.  The observed change in the quantity of liquid
provides a way of measuring $\Delta Q_{hot}$.  Likewise the cold
reservoir may consist of a substance in solid and liquid forms.  In
these examples, the temperatures of the hot and cold reservoirs do not
vary much even when significant amounts of heat are added to them or
removed.  The absolute temperature $T=T(\theta)$ is a monotonic
function of measured temperature $\theta$.  The relation between
absolute and measured temperatures may be found, e.g., in
\cite{schroeder} \cite{landau} \footnote{ From a practical
stand-point, absolute temperatures may be taken as proportional to the
volume of a gas such as helium at atmospheric pressure, except at very
low and very high temperatures.}.  Carnot considered that absolute
temperatures are obtained by adding 267 to thermometer readings
expressed in degrees Celsius, instead of the currently accepted value
of 273.15 \cite {car}, p.67, \cite {sadi}, p.  211.

Carnot first proved that engines attain their highest efficiencies
when they are \emph{reversible}.  To explain what ``reversible'' means,
let us suppose that the heat engine generates a work $\Delta W$, with
the hot bath loosing some amount of heat and the cold bath gaining
some.  If the engine is reversible the initial bath heat contents get
restored when the energy $\Delta W$ is fed in, in which case the
system is called a heat pump.  If the work performed by a reversible
heat engine of efficiency $\eta$ is employed to drive an identical
engine in the reversed mode, the heat engine-heat pump assembly does
not generate any net work.  There is no net heat consumption either,
so that the assembly may go on for ever, ideally.

It is not possible for a heat engine to have an efficiency greater
than the efficiency $\eta$ of reversible systems.  Indeed, such an
hypothetical heat engine operating with the same heat baths as before
and with the same heat consumption would generate a work exceeding
$\Delta W$.  If this heat engine were employed to drive the previously
considered heat pump, the hypothetical-heat-engine/heat-pump assembly
would perform positive work while the bath heat contents would remain
the same.  Energy would then be obtained for free, in violation of the
law of conservation of energy.  The above considerations apply of
course also to purely mechanical systems such as water-mills, whose
efficiency, ideally, is unity.  It is of historical interest that
water-mill reversibility was studied by Lazare Carnot (Sadi Carnot's
father).

Carnot employed a mechanical analogy.  Let us quote from his book
\cite{car}, on page 28: ``There is some justification in the
comparison between the motive power of heat and that of a
waterfall [\ldots] which depends on its height and the quantity of
liquid.  The motive power of heat depends also on the quantity of
entropy used and what one could designate [\ldots] as the height of its
fall, i.e., the difference of temperature between the bodies
exchanging entropy''.  We have translated ``calorique'' by entropy,
following the observation made by Zemansky \cite{zem}\footnote
{``Carnot used ``chaleur'' when referring to heat in general.  But
when referring to the motive power of heat that is brought about when
heat enters at high temperature and leaves at low temperature, he uses
the expression ``chute de calorique'', never ``chute de chaleur''.  It
is the opinion of a few scientists that Carnot had in the back of his
mind the concept of entropy, for which he reserved the term of
calorique.  This seems incredible, and yet it is a remarkable
circumstance that, if the expression ``chute de calorique'' is
translated fall of entropy, objections raised against Carnot's work
[\ldots] become unfounded''.  This quotation from Zemansky has been
slightly abbreviated for the sake of clarity.}.  In notes published
after his death in 1832, but probably written at the time his book was
being published, Carnot points out that heat is equivalent to
energy\footnote {``Heat is nothing but motive power or rather another
form of motion.  Wherever motive power is destroyed, heat is generated
in precise proportion to the quantity of motive power destroyed;
conversely, wherever heat is destroyed, motive power is generated''. 
Note that Carnot employs here the word ``chaleur'' (heat), not
``calorique'' (entropy).}, and calculates on the basis of imprecise
experimental observations that 1 calorie of heat is equivalent to 3.27
Joules of energy, instead of 4.184 Joules \cite{sadi}, p.  195.

\begin{figure}[p]
	\centering
	\includegraphics[scale=0.7]{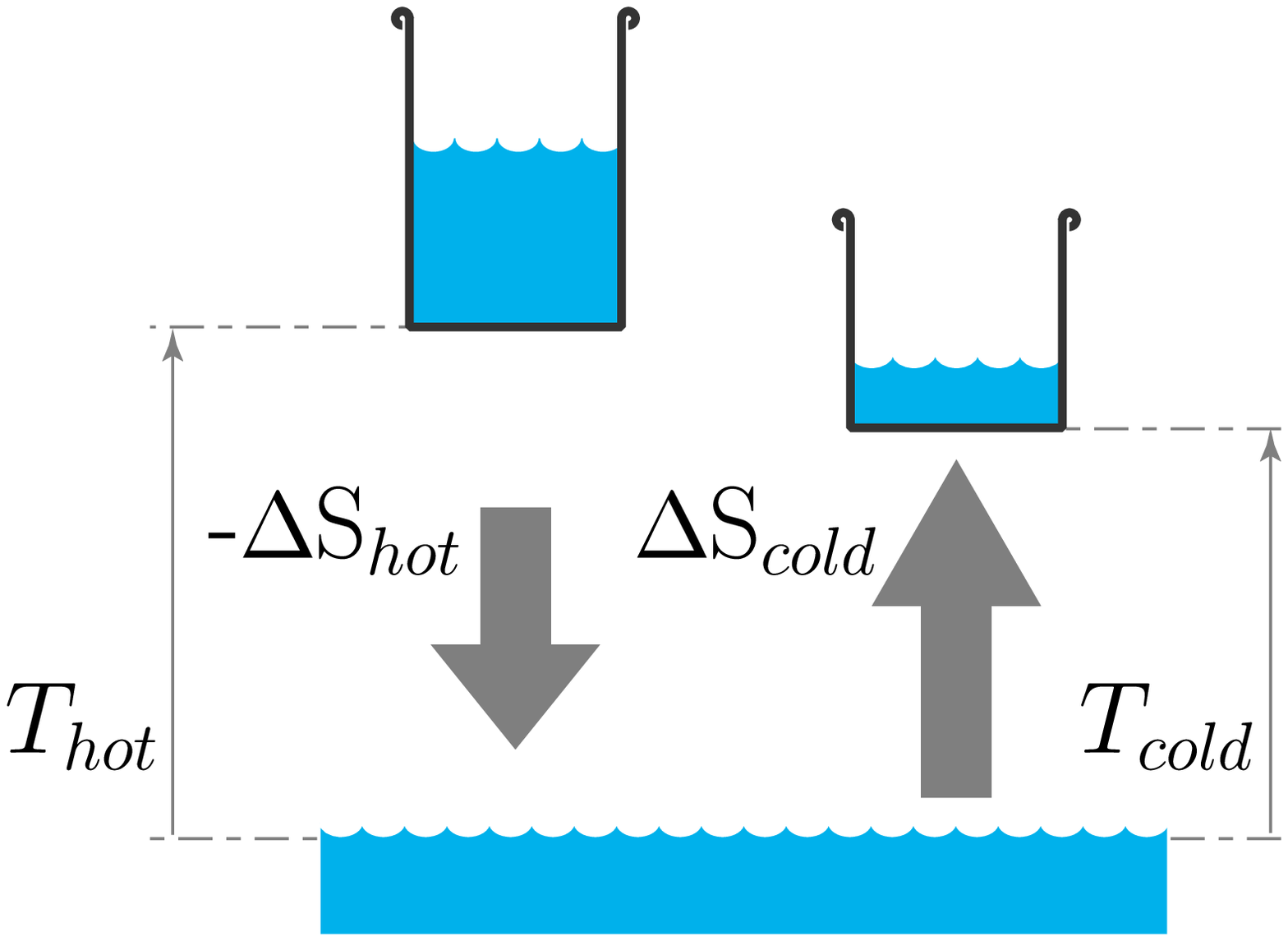}
	\caption{Water-fall picture of Carnot cycles for a system
	alternately in contact with hot and cold baths, pictured as
	reservoirs at altitudes $T_{hot}$ and $T_{cold}$,
	respectively, above a lake.  If the cycle is reversible, the
	amount of water flowing from the upper reservoir to the lake
	is equal to the amount of water pumped from the lake to the
	lower reservoir.  In that case the Carnot efficiency may be
	attained.  $\Delta \mathrm{S}_{hot}$ and $\Delta
	\mathrm{S}_{cold}$ represent the entropies produced in the two
	baths.}
	\label{fig:1}
\end{figure}

The Carnot analogy is illustrated in Fig.~\ref{fig:1}.  Consider a
reservoir at altitude $T_{hot}$ above a lake.  If some water weight
$-\Delta \mathrm{S}_{hot}$ flows from the reservoir to the lake the
work performed is $-\Delta
\mathrm{Q}_{hot}=-T_{hot}\Delta\mathrm{S}_{hot}$.  Consider another
reservoir at a lower altitude $T_{cold}$.  In order to pump a water
weight $\Delta\mathrm{S}_{cold}$ from the lake to the reservoir a work
$T_{cold}\Delta\mathrm{S}_{cold}$ is needed.  The net work performed
$\Delta
\mathrm{W}=-T_{hot}\Delta\mathrm{S}_{hot}-T_{cold}\Delta\mathrm{S}_{cold}$
may fluctuate from cycle to cycle.  The efficiency is defined as the
ratio of the average work performed and the average consumption of
heat from the hot bath: $\eta\equiv -\Delta W/\Delta
Q_{hot}=1+T_{cold}\Delta S_{cold}/T_{hot}\Delta S_{hot}$.  The
limiting Carnot efficiency quoted above obtains if $\Delta
S_{hot}+\Delta S_{cold}=0$, that is, if the average amount of water lost by
the upper reservoir ends up in the lower one.

To conclude that the efficiency may not exceed
$\eta_{C}=1-T_{cold}/T_{hot}$, one must prove that the total average
bath entropy produced never decreases, that is: $\Delta
S_{cold}+\Delta S_{hot}\ge 0$.  A concise argument is as follows: In
the case of heat pumps we have: $\Delta S_{hot}\ge 0$ and $\Delta
S_{cold}\le 0$.  Let us consider the special case for which
$T_{cold}\Delta S_{cold}+T_{hot}\Delta S_{hot}=0$, in which case no
work is involved ($\Delta W=0$).  This relation implies that $\Delta
S_{cold}+\Delta S_{hot}\le 0 $, since $T_{hot}\ge T_{cold}$, opposite
to the one we wish to prove.  But the situation just described does
not occur because heat never flows from cold to hot baths
spontaneously, according to observation.  The general result follows
from the fact that the temperatures may be specified arbitrarily.

To summarize: if the entropies produced in the hot and cold baths
$\Delta \mathrm{S}_{hot}= \Delta \mathrm{Q}_{hot} / T_{hot}$ and
$\Delta \mathrm{S}_{cold}= \Delta \mathrm{Q}_{cold} / T_{cold}$ can
be evaluated, the work performed by the engine
\begin{eqnarray}
\Delta \mathrm{W}=-\Delta \mathrm{Q}_{hot}-\Delta \mathrm{Q}_{cold}
	\label{workperformed}
\end{eqnarray}
is readily obtained since the bath temperatures are known.  This work
may fluctuate.  We are mostly interested in its average value $\Delta W$. 
The efficiency
\begin{eqnarray}
\eta=1+\frac{\Delta Q_{cold}}{\Delta Q_{hot}}=1+\frac {T_{cold}}{T_{hot}} 
\frac{\Delta S_{cold}}{\Delta S_{hot}}
	\label{rendement}
\end{eqnarray}
reaches the Carnot efficiency when the cycle is reversible, that 
is, when the total average bath entropy 
produced per cycle $\Delta S_{hot}+\Delta S_{cold}$ vanishes.

Two seemingly independent entities were considered above, namely the
absolute temperature $T$ of a bath, analogous to a reservoir height,
and a state function $\mathrm {S}$, called entropy, analogous to the
total weight of water contained in a reservoir.  To proceed further,
we need introduce another state function, namely the energy
$\mathrm{U}_{bath}$ contained in the bath.  It is plausible that
$\mathrm{S}$ be some function of $\mathrm{U}_{bath}$, since both are
state functions and no other parameter is presently involved.  If an
amount of heat $\Delta \mathrm{Q}\ll \mathrm{U}_{bath} $ is added, the
bath energy gets incremented by $\Delta\mathrm{U}_{bath}=\Delta
\mathrm{Q}$, according to the law of equivalence of heat and energy. 
The bath entropy gets incremented by $\Delta\mathrm{S}=\beta~ \Delta
\mathrm{Q}=\beta ~\Delta\mathrm{U}_{bath}$, if the inverse bath
temperature $\beta\equiv 1/T$ is introduced.

Carnot explained how reversible heat engines could be constructed: He
first observed that two bodies should be put into thermal contact only
if their temperatures almost coincide.  Reversible transformations
must be quasi-static, that is, close to an equilibrium state at every
instant.  As a consequence, ideal heat engines, while efficient, are
slow.  Reversible heat engines involve four steps, two of them with
the system being isolated from the baths (adiabatic transformations),
and two of them with the system being in contact with either the hot
or the cold baths (isothermal transformations).  These four steps will
be discussed in detail in subsequent sections.
 
In the 19th century only systems involving many microscopic degrees of
freedom, such as gas-filled cylinders terminated by movable pistons,
were considered.  We discuss here instead single-mode oscillators that
possess a single degree of freedom, the phase of the oscillation being
ignored. The conditions under which the cycle should be
considered reversible will need clarification.  Carnot cycles
involving oscillators were discussed before (see \cite{sekimoto}, and
the references therein).  Small mechanical systems have been
considered, e.g., in \cite{serry}, and the Statistical Mechanical
properties of small electronic systems are discussed, e.g., in
\cite{arnaud}.  The forces involved in single-mode oscillators are
tiny.  But rotating-vibrating molecules and biological systems
submitted to baths at different temperatures may retrieve energy
through Carnot cycles or related devices \cite {reimann}.

The general expression of the work performed by a system in contact
with a bath when its parameter varies \cite{kittel} is recalled in
Section \ref{2}.  It is shown in Section 3 that for linear oscillators
at frequency $\omega$ the Carnot result amounts to asserting that the
average oscillator action, $f(x)$, is a decreasing function of $x
\equiv \beta \omega$.  The properties of Carnot cycles for oscillators
are discussed in Section \ref{4}.  The explicit form of $f(x)$ is
obtained in Section \ref{5} from a simple prescription.  The average
work $\Delta W$ performed per cycle and the efficiency $\eta$ are
illustrated for reversible and non-reversible cycles in Section
\ref{6}.

Szilard noted in 1925 that: ``exploitation of the fluctuation
phenomena will not lead to the construction of a perpetual mobile of
the second kind'' \cite{szilard}.  It may be shown on the basis of the
Boltzmann formulation that the variance as well as the average value
of the entropy produced vanish when the reversibility conditions are
fulfilled, in agreement with that quotation.

The Boltzmann constant $\kB$, set equal to unity for brevity, is
restored in numerical applications.  The angular frequency $\omega$ is
called ``frequency'' for short, the Planck constant divided by $2\pi$,
$\hbar$, is called ``Planck constant'', and the action divided by
$2\pi$, namely $\mathrm{f}=\mathrm{U}/\omega$, is called ``action''.

\section {Work performed by a system in contact with a bath}
\label{2}

Let a system depending on a parameter $\omega$ (perhaps the system
volume) interact weakly with a bath.  Some energy flows between the
bath and the system. Eventually a state of equilibrium is
reached.  The system average energy, denoted $U(\omega, \beta)$,
depends on $\omega$ and on the bath inverse temperature $\beta $.

If the parameter varies by $d\omega$, the elementary average work $dW$
performed by the system is written as $dW=-f \, d\omega$, where $f$ is
a generalized force\footnote{The notations $dW$ or $dQ$ are \emph{not}
meant to imply that these quantities are total differentials.  }.  If
$\omega$ represents the volume of a gas-filled enclosure, $-f$
represents the gas pressure.  If $\omega$ represents an oscillator
frequency, $f$ represents the oscillator \emph{action}.  Purely
mechanical considerations, such as the law of conservation of
momentum, often enable one to evaluate the generalized force as some
function $f(\omega ,U)$ of $\omega $ and the average system energy
$U$.  Considering $U$ as a function of $\omega$ and temperature
reciprocal $\beta$, as said above, we may view $f$ as a function of
$\omega$ and $\beta $ according to $f(\omega,\beta )\equiv
f(\omega,U(\omega,\beta ))$.  In the present paper the parameter
$\omega$ is supposed to be prescribed from the outside, that is, it is
not subjected to fluctuations.  Bath temperatures are of course fixed
quantities.

The fact that the bath entropy $S$ is a state function restricts
admissible functions $U(\omega ,\beta )$.  Indeed, the bath entropy
increment $dS=\beta dQ=-\beta(dW+dU)\equiv \beta (fd\omega -dU)$, if
the law of conservation of energy ($dQ+dW+dU=0$) and the definition of
$f$ are introduced.  $dS$ may be expressed as
\begin{eqnarray}
	dS = \beta (f-\frac{\partial U}{\partial \omega}) d\omega -
	\beta \, \frac{\partial U}{\partial \beta } \, d\beta.
	\label{dSS}
\end{eqnarray}
Because $dS$ is a total differential, the derivative with respect to
$\omega $ of the term that multiplies $d\beta $ must be equal to the
derivative with respect to $\beta $ of the term that multiplies
$d\omega $.  After simplification, we obtain
\begin{equation}\label{dddS}	
	\frac{\partial U}{\partial \omega} = \frac{\partial (\beta
	f)}{\partial \beta} .
\end{equation} 

Since $dW=-f~d\omega$, the work $\Delta W_{isothermal}$ performed by
the system in contact with the bath when the parameter varies from
$\omega_{in}$ to $\omega_{out}$ is given by
\begin{eqnarray}\label{isothermal}	
	\beta \Delta W_{isothermal}(\beta)=-\beta
	\int_{\omega_{in}}^{\omega_{out}}f(\omega,\beta)d\omega
	=\phi(\omega_{in},\beta)-\phi(\omega_{out},\beta).
\end{eqnarray}
where we have defined
\begin{eqnarray}\label{freex}	
	 \phi(\omega,\beta)\equiv \beta \int^\omega f(\omega',\beta)
	 d\omega'.
\end{eqnarray}

The evolution of a system in contact with a bath may be pictured as a
sequence of elementary adiabatic evolutions with $\omega$ being
incremented by $d \omega$, followed by returns to thermal equilibrium
at constant $\omega$.  Because the successive steps are statistically
independent, the average values and variances of the elementary works
performed during the adiabatic steps add up.  It follows that the
ratio of the standard deviation (square root of the variance) to the
average work goes to zero as $1/\sqrt {N}$ when the number $N$ of
elementary steps increases.  In other words, the work produced may be
considered as a non-fluctuating quantity provided the process is
sufficiently slow.

\section{Linear oscillators}
\label{3}

For concreteness, let us consider an inductance-capacitance
$\mathcal{L}-\mathcal{C}$ circuit resonating at angular frequency
$\omega$, as shown in Figs.~\ref{fig:2}a and \ref{fig:2}b.  The
electrical charges on the capacitor plates oscillate sinusoidally in
the course of time, but the two plates always attract each others.  It
follows from the Coulomb law that the cycle average force $\mathcal{F}
= \mathrm{U} / 2 a$, where $\mathrm{U}$ denotes the resonator energy
and $a$ the capacitor plates separation.  If $a$ is incremented by
$da$ slowly so that the oscillation remains almost sinusoidal, the
elementary work performed by the oscillator is $d\mathrm{W} =
-\mathcal{F} \, da = \mathrm{U} \, da / 2 a$.  On the other hand, it
follows from the well-known resonance condition $\mathcal{L} \,
\mathcal{C} \, \omega^2 = 1$ and the fact that $\mathcal{C}\propto
1/a$, where $\propto $ denotes proportionality, that $2 \,
d\omega/\omega = da/a$.  The elementary work may therefore be written
as $d\mathrm{W} = -(\mathrm{U}/\omega)~d\omega \equiv -\mathrm {f}~
d\omega$, where we have introduced the generalized force $\mathrm
{f}=\mathrm{U}/\omega$.  For the average values we have $f=U/\omega$.

When the system is isolated, i.e., \emph{not} in contact with a heat
bath, we have $d \mathrm{Q} = 0$ and thus $d\mathrm{U}
+d\mathrm{W}=0$.  According to the previous expression of
$d\mathrm{W}$ the oscillator energy gets incremented by $d\mathrm{U} =
(\mathrm{U}/\omega) \, d\omega$.  It follows that when the resonant
frequency of an isolated oscillator varies slowly, the ratio
$\mathrm{U}/\omega $, called ``action'', does not vary significantly. 
In other words, the generalized force $\mathrm{f}=\mathrm{U}/\omega $
is constant in adiabatic processes in the case of oscillators\footnote
{More formally, the Hamiltonian $H(q,p,t)=\frac{1}{2}[p^2+\omega^2(t)
q^2]$ for a non-relativistic particle of mass $m=1$ in a potential
well $V(x,t) = \omega^2(t) x^2 / 2$, and $dq/dt = p$, $dp/dt =
-\omega^2(t) q$.  A straightforward derivation shows that
$\frac{d}{dt}[H(q(t),p(t),t)/\omega(t)]\approx 0$, if we take into
account the fact that the average kinetic energy is equal to the
average potential energy: $<p^2-\omega^2 q^2>=0$.}.

Replacing $U$ in (\ref {dddS}) by $\omega f$, we obtain after
simplification
\begin{equation}\label{partialf}	
	\omega \, \frac{\partial f}{\partial \omega } = \beta \,
	\frac{\partial f}{\partial \beta} ,
\end{equation} 
a relation that entails that $f$ is a function of
$\ln(\omega)+\ln(\beta)=\ln(\beta\omega)$ only.  This is essentially
the displacement law discovered by Wien in 1893 \cite{wien}: blackbody
spectra scale in frequency in proportion to temperature\,\footnote{The
average oscillator energy $U$ is equal to $\omega f(\beta \omega)$,
where $f(x)$ is a function of a single variable to be later specified. 
The blackbody radiation spectral density is obtained by multiplying
$U$ by the electromagnetic mode density.  In the case of a cavity of
large volume $V$, a mode count shows that the number of modes whose
frequency is comprised between $\omega$ and $\omega+d\omega$ is equal
to $V\omega^2~d\omega/\pi^2~c^3$.  Thus the radiation spectral density
is proportional to $V\omega^3f(\omega/T)$.  If $T$ is multiplied by
some constant $a$, the spectrum therefore needs only be rescaled
frequency-wise by the same factor $a$.  Provided the integral
converges, the total black-body radiation energy density $u$, obtained
by integrating the previous expression over frequency and dividing by
$V$, reads $u=\sigma T^4$, where the Stefan-Boltzmann constant
$\sigma$ obtains from measurement.

Historically, the Wien displacement law obtained through quite a
different route.  First, Kirchhoff in 1860 established that the
radiation energy density $u$ in a cavity of large volume $V$ is a
function of temperature $T$ only.  Maxwell proved that the pressure
exerted by a plane wave on a perfectly reflecting mirror is equal to
the wave energy density $u$, a result which, incidentally, holds for
any isotropic non-dispersive wave, see, e.g., \cite{arnaud2}.  If we
take into account the fact that the direction of the incident wave is
randomly and uniformly distributed, the radiation pressure reads in
three dimensions: $\mathcal{P}=u/3$.  Boltzmann employed the laws of
Thermodynamics and established that $u\propto T^4$.  This conclusion
readily follows from (\ref{dddS}) of the present paper with the
substitutions: $\omega \to V$, $f\to -\mathcal{P}=-u(T)/3$.  Finally,
Wien in 1893 observed that the light reflected from a slowly moving
piston is frequency shifted, and enforced conditions for the radiation
spectrum to be at equilibrium.  The Wien reasoning is notoriously
difficult.  Interested readers will find the details in \cite
{darrigol}.  For a multimode treatment see, e.g., \cite{cole}.}.

Since $f$ is a function of $\beta\omega\equiv x$ only, the
$\phi$-function defined in (\ref{freex}) may be written as
\begin{equation}	
	\phi(\omega , \beta) = \beta \int^\omega f(\omega',\beta) \,
	d\omega'= \int^x f(x') \, dx'\equiv \phi(x).
	\label{psi'}
\end{equation}

\section {The Carnot cycle}
\label{4} 

Let us first consider the adiabatic processes.  Let $\mathrm{U}_{1}$
denote the oscillator energy when it is separated from the hot bath. 
If the frequency is changed slowly from $\omega_{1}$ to $\omega_{2}$,
we have
\begin{equation}\label{isol1}	
	\mathrm{U}_{2}=\frac{\omega_{2}}{\omega_{1}}\mathrm{U}_{1}\equiv
	\omega_{2}\mathrm{f}_{1},
\end{equation}
since isolated oscillator
energies are proportional to frequency. Likewise,
\begin{equation}\label{isol2}	
	\mathrm{U}_{4}=\frac{\omega_{4}}{\omega_{3}}\mathrm{U}_{3}\equiv
	\omega_{4}\mathrm{f}_{3}.
\end{equation}

Using the results (\ref{isothermal})-(\ref{freex}),
(\ref{psi'})-(\ref{isol2}), the entropies produced in the hot and cold
baths read respectively
\begin{eqnarray}\label{SSS1}	
	\beta_{hot} \Delta\mathrm{Q}_{hot} & = & \beta_{hot}
	(\mathrm{U}_{4} - \mathrm{U}_{1} -
	W_{isothermal}(\beta_{hot})) \nonumber \\
	&=&(a_{4}\mathrm{f}_{3} - \phi(a_{4})) - (a_{1}\mathrm{f}_{1}
	- \phi(a_{1}))
\end{eqnarray}
and
\begin{eqnarray}\label{SSS2}	
	\beta_{cold} \Delta\mathrm{Q}_{cold} & =
	&\beta_{cold}(\mathrm{U}_{2} - \mathrm{U}_{3} -
	W_{isothermal}(\beta_{cold}))\nonumber\\
	&=& (a_{2}\mathrm{f}_{1} - \phi(a_{2})) - (a_{3}\mathrm{f}_{3}
	- \phi(a_{3})),
\end{eqnarray}
where we have defined 
\begin{align}
	a_{1} & \equiv \beta_{hot} \omega_{1} ,
	& a_{2} & \equiv \beta_{cold} \omega_{2} ,
	& a_{3} & \equiv \beta_{cold} \omega_{3} ,
	& a_{4} & \equiv \beta_{hot} \omega_{4} .
	\label{xy}
\end{align}
The above expressions of the $\Delta\mathrm{Q}$ follow simply from the 
law of conservation of energy. Recall that in these expressions 
$W_{isothermal}$ is a non-fluctuating quantity.

The average entropies produced are obtained by replacing in
(\ref{SSS1}) and (\ref{SSS2}) $\mathrm{f}_{1}$ by $f(a_{1})$ and
$\mathrm{f}_{3}$ by $f(a_{3})$:
\begin{eqnarray}	
	\beta _{hot}\Delta  Q_{hot}& = &  s(a_{4},a_{3})-s(a_{1}) \nonumber\\
	\beta _{cold}\Delta Q_{cold}& = &  s(a_{2},a_{1})-s(a_{3})  ,
	\label{work'}
\end{eqnarray}
where we have introduced a function of two variables 
\begin{eqnarray}	
	s(x,y)  \equiv  x \, f(y) - \phi(x) 
	 \equiv  x \, f(y) - \int^x f(x') \, dx', 
	\label{sxyz}
\end{eqnarray}
and $s(x)\equiv s(x,x)$.  Note for later use that the function
$s(x,y)$ is unaffected by the addition of a constant to $f$.  We
further observe that $s(x,y) - s(y)\ge 0$ if $f$ is a decreasing
function of its argument.  We will later on show that this condition,
equivalent to the Carnot principle, indeed holds.  The average work
performed by the system per cycle and the efficiency follow from
(\ref{work'}) according to the general expression
(\ref{workperformed}) and (\ref{rendement}) if the function $f(x)$ is
known.  The Carnot efficiency is attained when $a_{1}=a_{2}$ and
$a_{3}=a_{4}$, that is, when the reversibility conditions
\begin{equation}\label{revers}	
	\frac{T_{cold}}{T_{hot}} = \frac{\omega _{2}}{\omega_{1}} =
	\frac{\omega _{3}}{\omega _{4}}
\end{equation} 
hold.  It is interesting
that this condition is independent of the form of the $f(x)$
function. 

The total average bath entropy produced per cycle is
\begin{eqnarray}\label{totS}	
	\Delta S_{t}&=&\beta_{cold}Q_{cold}+\beta_{hot}Q_{hot}= 
	s(a_{2},a_{1})-s(a_{1}) +s(a_{4},a_{3})-s(a_{3})
\end{eqnarray}
This quantity is non-negative if $f(x)$ is a decreasing function of 
$x$, according to a previous remark.
For small departures from reversibility, i.e., for $a_{1}\approx
a_{2}, a_{3}\approx a_{4}$, expansion up to second order of the above 
expression gives
\begin{eqnarray}\label{approxS}	
	\Delta S_{t}\approx -\frac{1}{2}(a_{2}-a_{1})^2 
	\frac{df(a_{1})}{da_{1}}-\frac{1}{2}(a_{4}-a_{3})^2 
	\frac{df(a_{3})}{da_{3}}
\end{eqnarray}

The Boltzmann-Gibbs formulation tells us that the probability that the
system energy be $\epsilon_{k}(\omega)$, $k=0,1\ldots$, is
proportional to $\exp(-\beta \epsilon_{k}(\omega))$.  One can prove
from that formulation that $-df/dx$ is equal to the variance of the
oscillator action and is therefore positive.  This relation shows
further that the variance of the total entropy produced per cycle is
twice the average value given in (\ref {approxS}), a conclusion
related to the ones given in \cite{crooks} and \cite{jarzynski}. 
These two papers consider only classical systems in contact with a
bath, but they are much more general on other respects.

Furthermore, it can be shown that cycles are reversible if and only if
\begin{equation}\label{reversbis}	
	\frac{T_{cold}}{T_{hot}} = \frac{ \epsilon_{k}( \omega _{2} )
	- \epsilon_{0}( \omega_{2} )}{\epsilon_{k}( \omega_{1} ) -
	\epsilon_{0}( \omega_{1}) } = \frac{\epsilon_{k}( \omega_{3} )
	- \epsilon_{0}( \omega_{3} )}{\epsilon_{k}( \omega_{4} ) -
	\epsilon_{0}( \omega_{4} )},
\end{equation} 
for $k=1,2,\ldots$.  These relations may hold when the
$\epsilon_{k}(\omega)$ factorize as $h(k)g(\omega)$.  For oscillators
in particular we have $\epsilon_{k}(\omega)=(k+1/2)\hbar \omega$,
according to the Quantum Optics formulation, and the simpler
expression in (\ref{revers}) is recovered.  The above result is based
on the concept of relative entropy \cite{qian}.  The mathematical
details will be given elsewhere.

\section{Average oscillator energy}
\label{5}

According to the Boltzmann-Gibbs formulation, the average energy of a
classical one-dimensional oscillator is $U=T$.  Thus, the average
oscillator action $f=1/\beta\omega \equiv 1/x $ obeys the differential
equation
\begin{equation}\label{diff}	
	\frac{df}{dx} + f^2 = 0.
\end{equation}
The relation $U=T$, however, is unacceptable because it would
necessarily lead to infinite blackbody radiation energy if the Maxwell
electromagnetic theory is to be upheld.  Indeed, the Maxwell theory
applied to a cavity having perfectly conducting walls predicts that
there is an infinite number of modes, each of them being modeled as an
harmonic oscillator.  If an average energy $T$ is ascribed to them,
the total energy is clearly infinite.  This observation, made near the
end of the 19th century, caused a major crisis in Physics \cite{kuh}. 
It apparently did not occur to the physicists facing that problem that
the mere addition of a constant on the right-hand-side of (\ref{diff})
would solve the problem.  Let us indeed suppose that
\begin{equation}\label{difff}	
	\frac{df}{dx} + f^2 = \left( \frac{\hbar}{2} \right)^2 .
\end{equation}
where $\hbar$ is now known as the Planck constant.  Note that $f$ and
$\hbar$ have the dimension of action (`energy' $\times$ `time'), while
$x$ has the dimension of an action reciprocal, so that $\hbar x$ is
dimensionless.

The solution of 
(\ref{difff}) that gives $f(x)-1/x\to 0$ in the classical limit $T\to 
\infty$ reads
\begin{eqnarray}	
	f(x) = \frac{\hbar}{2}+\frac{\hbar}{\exp(\hbar x) - 1} .
	\label{fff}
\end{eqnarray}
This is the expression obtained by Planck in 1900 from a fit to the
available experimental results, aside from the term $\hbar/2$.  The
latter term is responsible for the Casimir effect \cite{cooke},
\cite{rev} but, as we have seen, it does not affect cyclic operations. 
We obtain from (\ref{fff}) after integration
\begin{equation}\label{ff}	
	\phi(x) = \frac{\hbar x}{2} + \ln \left[
	1-\exp(-\hbar x) \right] ,
\end{equation}
Therefore, the entropic function defined in (\ref{sxyz}) reads 
\begin{equation}\label{sxy}	
	s(x,y) = \frac{\hbar x}{\exp(\hbar y)-1} - \ln \left[
	1-\exp(-\hbar x) \right] .
\end{equation}
When this result is introduced in (\ref{work'}), explicit expressions
for the work performed (\ref{workperformed}) and efficiency
(\ref{rendement}) follow.  In the classical regime, $T\to \infty$, the
above expression reduces to
\begin{equation}\label{sxybis}	
	s(x,y) = \frac{ x}{y} - \ln(\hbar x) .
\end{equation}
Note that the Planck constant $\hbar$ cancels out in the final
formulas.  We leave it there for aesthetic reasons, the argument of
$\ln(.)$ being expected to be dimensionless.

The quantities of interest in a Carnot cycle, i.e., mainly the work
performed and the efficiency, have been obtained without giving any
consideration to the oscillator entropy or temperature.  We only need
evaluate the entropies produced in the cold and hot baths.  For
purposes of illustration (see Fig.~\ref{fig:2}) it is, however, of
some interest to introduce the oscillator inverse temperature
$\beta_{osc.}$.  When the oscillator is in contact with a bath at
inverse temperature $\beta$ we have $\beta_{osc.}=\beta $.  When the
oscillator is isolated and the frequency varies slowly the product
$x=\beta_{osc.} \omega$ remains constant, as we have seen.  The
average system entropy is a function of $x\equiv \beta_{osc.}\omega$
only, and thus remains constant during the adiabatic process.  The
condition $a_{1}=a_{2}$ stated above amounts to saying that
reversibility requires that the oscillator be put into contact with
the cold bath only if its inverse temperature is almost equal to
$\beta_{cold}$, and likewise for the other adiabatic process.  Let us
emphasize, however, that this simple picture, similar to the one
advanced by Carnot in 1824, does not generally apply to multimode
oscillators, unless the modes frequencies vary in proportion of one
another, or are continuously thermalized through some non-linear
coupling.

\section  {Illustration of Carnot cycles.}
\label{6}

Let us give first an order of magnitude of the work performed,
considering for simplicity the classical limit: $T\to \infty $.  In
that limit: $s(x) = -\ln(x) + \mathrm{constant}$.  It follows that for
reversible classical cycles the average work performed per cycle reads
\begin{equation}\label{cl}	
	\Delta W = \kB (T_{hot}-T_{cold}) \ln \left(
	\frac{\omega_{3}}{\omega_{1}} \frac{T_{hot}}{T_{cold}} \right),
\end{equation}
where the Boltzmann constant has been restored.  At room temperature
the classical approximation is a valid one if the oscillator
frequencies are substantially smaller than about 10~THz.  For example,
for $N = 1/\kB \approx 10^{23}$ independent oscillators,
$T_{hot}=1200$~K, $T_{cold}=300$~K, and $\omega_{1} = 2 \omega_{3}$,
the work done per cycle $\Delta W=900 \ln(2)$~J $\approx 620$~J.

\begin{figure}[p]
	\centering
	\includegraphics[scale=0.7]{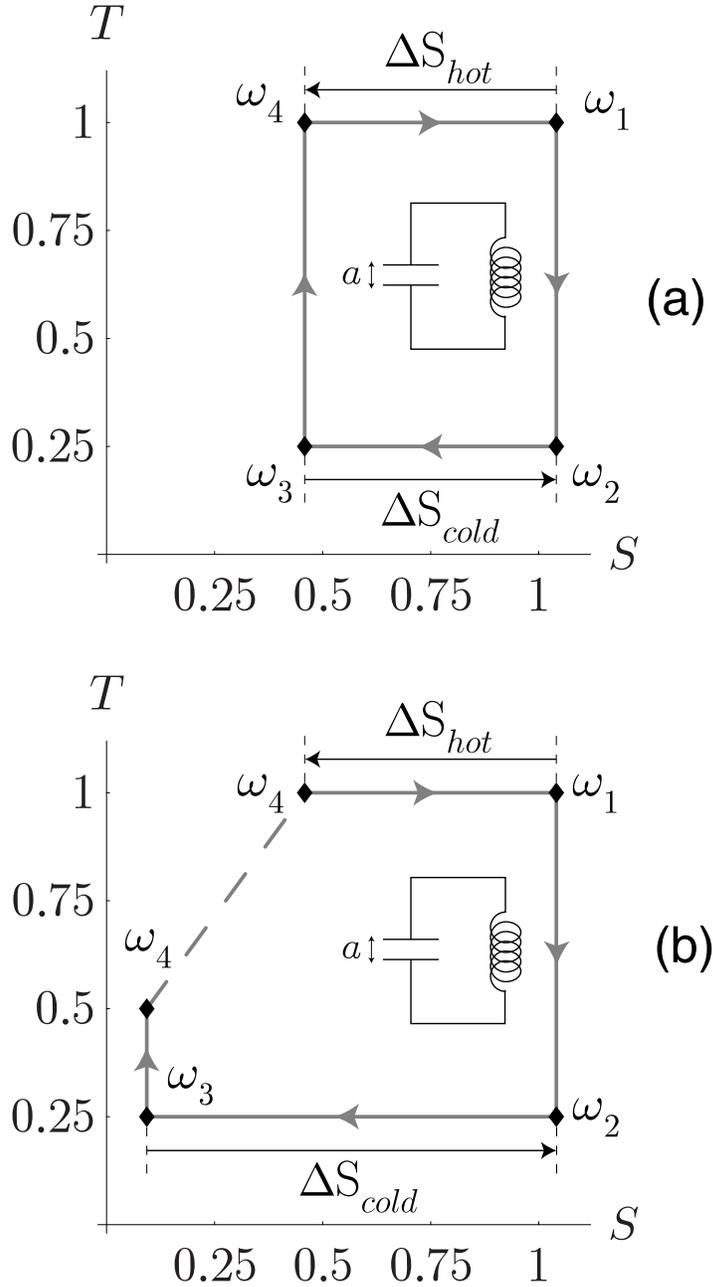}
	\caption{An $\mathcal{L} - \mathcal{C}$ single-mode oscillator
	is shown.  The figure shows cycles in the temperature versus
	oscillator entropy diagram for the case where $T_{hot}=1$ and
	$T_{cold}=1/4$, $\omega_{1}=1$, $\omega_{2}=1/4$,
	$\omega_{3}=c/2$, $\omega_{4}=2$.  (a) Reversible cycle,
	$c$=1.  (b) Non-reversible cycle, $c$=2.  The horizontal
	arrows give the average entropies \emph {produced} in the cold
	(lower) and hot (upper) baths.}
	\label{fig:2}
\end{figure}

Let us now go back to the quantum regime and evaluate explicitly the
work performed and the efficiency of oscillators alternately in
contact with a hot bath at temperature $T_{hot}=1$ and a cold bath at
temperature $T_{cold}=1/4$.  We consider in Fig.~\ref{fig:2} the case
where $\omega_{1}=1$, $\omega_{2}=1/4$, $\omega_{4}=2$ and
$\omega_{3}$ is kept as a parameter.  When $\omega_{3}=1/2$, the
system is reversible and the cycle in the oscillator
temperature-versus-entropy diagram in Fig.~\ref{fig:2}a is
rectangular.  The entropy $\Delta S_{cold}$ produced in the cold bath
is shown by the lower right-directed arrow, while the entropy $-\Delta
S_{hot}$ removed from the hot bath is shown by the upper,
left-directed arrow.  The total produced entropy vanishes in that
case.  The case of an irreversible cycle with $\omega_{3}=1$ is shown
in Fig.~\ref{fig:2}b.  Note the temperature-entropy jump, shown by a
dashed line, when the oscillator is put in contact with the hot bath. 
In that situation the total produced entropy (see the arrows) is
positive.

\begin{figure}[p]
	\centering
	\includegraphics[scale=0.7]{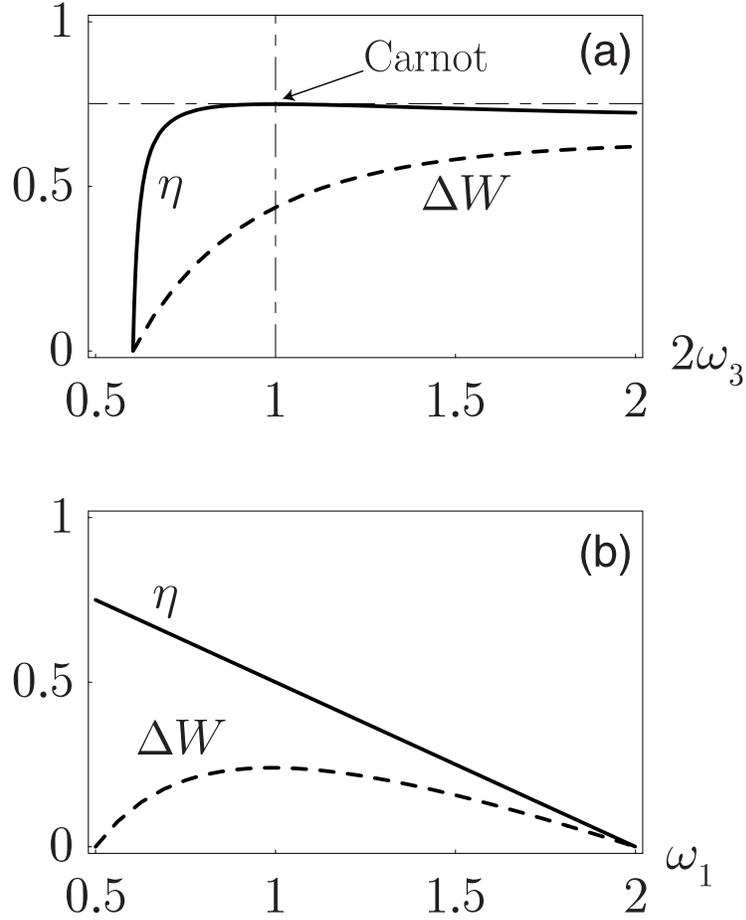}
	\caption{Average work performed per cycle $\Delta W$ and
	efficiency $\eta$ for $T_{hot}=1$ and $T_{cold}=1/4$, (a) as
	functions of $2\omega_{3}$.  The Carnot efficiency is attained
	when $2\omega_{3}=1$.  (b) The oscillator frequency is kept
	constant when in contact with either bath.  We have set:
	$\omega_{2}=\omega_{3}=1/4$.  $\omega_{1}=\omega_{4}$ is kept
	as a parameter.  }
	\label{fig:3}
\end{figure}

Figure~\ref{fig:3}a shows how the work performed $\Delta W$ and the
efficiency $\eta$ vary as a function of $\omega_{3}$.  The Carnot
efficiency $\eta_{C}=3/4$ is reached for $\omega_{3}=1/2$.  For larger
$\omega_{3}$-values the energy extracted per cycle increases but the
efficiency is somewhat reduced.

A case of interest is when the resonator frequency is a constant
$\omega_{1}$ when the resonator is in contact with the hot bath and a
constant $\omega_{2}$ when it is in contact with the cold bath, in
which case $a_{1}=a_{4}$ and $a_{2}=a_{3}$.  In that situation, the
hot and cold baths may be modeled as large collections of oscillators
at frequencies close to $\omega_{1}$ and $\omega_{2}$, respectively. 
Supposing again that $T_{hot}=1$ and $T_{cold}=1/4$.  We find from
previous expressions that the energy extracted per cycle is maximum
when $\omega_{1}=1/2$ and $\omega_{2}=1/4$.  For these parameter
values the efficiency is $\eta=1/2$, that is, substantially less than
the Carnot efficiency $\eta_{C}=3/4$.  The variations of the work done
and the efficiency as functions of $\omega_{1}$ are shown in
Fig.~\ref{fig:3}b.

Another case of interest is when the parameter does not vary when the 
system is transferred from one bath to another. In that case, 
according to the observation that follows (\ref {freex}), the work 
performed does not fluctuate.

\section{Conclusion}

We have shown that heat engines whose system is a single-mode linear
oscillator obey the Carnot theory.  Explicit expressions for the work
performed per cycle and the efficiency were obtained on the basis of a
simple prescription.  We have illustrated reversible and
non-reversible cycles, and shown that the variance of the entropy
production per cycle vanishes when the cycle is reversible and is, in 
general, equal to
twice the average value.

The present theory may be generalized to multimode oscillators by
adding up modal contributions.  Consider in particular a
non-dispersive transmission line terminated by a movable
short-circuit, a configuration resembling the classical gas-filled
cylinder with a piston.  Because the resonant frequencies change in
proportion to one another when the length of the transmission line is
modified, a temperature may be defined at every step of the adiabatic
process and the Carnot efficiency may be attained.  This is also the
case when the shape of a cavity does not change as the volume varies. 
But this is not so for dispersive transmission lines such as
waveguides.  Slow length changes create an average distribution among
the modes that cannot be described by a temperature, unless some
thermalization mechanism is enforced at each elementary step.  Carnot
cycles for radiation are discussed in \cite{lee}.

The force $f$ depends on the term $\hbar \omega / 2$ in the expression
of the mode average energy, and is non-zero even at $T=0$K. But if we
are only interested in the average work performed over a full cycle,
this term may be ignored.

Recent interesting generalizations take into account finite
interaction ti\-mes, $\Delta t$, between the oscillator and the baths. 
In that case there are departures of the work done from the change in
free energy, which are inversely proportional to $\Delta t$.  Note
also that the energy required to detach a system from a bath should be
accounted for when the cycle is not slow \cite{sekimoto}.

\section{Acknowledgments}

The authors wish to express their thanks to E. Clot and J.C. Giuntini
for a critical reading of the manuscript.

\end{document}